\documentclass[10pt,conference]{IEEEtran}
\usepackage[utf8]{inputenc}
\usepackage[T1]{fontenc}
\usepackage{graphicx}
\usepackage{multirow}
\usepackage{booktabs}
\usepackage{xcolor}
\usepackage{dirtytalk}
\usepackage{comment}
\usepackage{hyperref}
\usepackage{amssymb}
\usepackage{caption}
\usepackage{graphicx}
\usepackage{subcaption}
\usepackage{adjustbox}
\usepackage{balance}
\usepackage[tikz]{mdframed}
\usepackage{showexpl}
\usepackage{enumitem}

\title{On the Prevalence, Evolution, and Impact of Code Smells in Simulation Modelling Software}

\author{
\author{
\IEEEauthorblockN{Riasat Mahbub}
\IEEEauthorblockA{\textit{Faculty of Computer Science} \\
\textit{Dalhousie University}\\
Nova Scotia, Canada \\
rs955802@dal.ca}
\and
\IEEEauthorblockN{Mohammad Masudur Rahman}
\IEEEauthorblockA{\textit{Faculty of Computer Science} \\
\textit{Dalhousie University}\\
Nova Scotia, Canada \\
masud.rahman@dal.ca}
\and
\IEEEauthorblockN{Muhammad Ahsanul Habib}
\IEEEauthorblockA{\textit{Department of Civil and Resource Engineering} \\
\textit{Dalhousie University}\\
Nova Scotia, Canada \\
ahsan.habib@dal.ca}
}
}

\begin{document}

\maketitle

\begin{abstract}
Simulation modelling systems are routinely used to test or understand real-world scenarios in a controlled setting. They have found numerous applications in scientific research, engineering, and industrial operations. Due to their complex nature, the simulation systems could suffer from various code quality issues and technical debt. However, to date, there has not been any investigation into their code quality issues (e.g. code smells). In this paper, we conduct an empirical study investigating the prevalence, evolution, and impact of code smells in simulation software systems. First, we employ static analysis tools (e.g. Designite) to detect and quantify the prevalence of various code smells in 155 simulation and 327 traditional projects from Github. Our findings reveal that certain code smells (e.g. Long Statement, Magic Number) are more prevalent in simulation software systems than in traditional software systems. Second, we analyze the evolution of these code smells across multiple project versions and investigate their chances of survival. Our experiments show that some code smells such as Magic Number and Long Parameter List can survive a long time in simulation software systems. Finally, we examine any association between software bugs and code smells. Our experiments show that although Design and Architecture code smells are introduced simultaneously with bugs, there is no significant association between code smells and bugs in simulation systems.
\end{abstract}

\section{Introduction}
\label{sec:intro}
Code smells are symptoms that indicate deeper code quality issues in software systems \cite{fowler2018refactoring}. They do not prevent a software program from working correctly but increase its technical debt over time. They could also lead to performance issues \cite{10.1145/2897073.2897100}, lack of understandability \cite{5741260}, and lack of maintainability in software systems \cite{6606614,6405287}. According to a study by Jaafar et al. \cite{jaafar2017analysis}, Object-Oriented classes with anti-patterns and code clones are three times more likely to contain faults or bugs than non-smelly classes. Moreover, Hecht et al. \cite{10.1145/2897073.2897100} suggest that refactoring of code smells can lead to 3.6\% memory and 12.4\% UI performance improvements in Android applications. Thus, the study of code smells has been an active research topic in the software engineering community \cite{SHARMA2018158,agnihotri2020systematic,lacerda2020code}.

Over the last two decades, there have been numerous studies on the prevalence  \cite{cardozo2023prevalence,saboia2020prevalence,8813267}, evolution \cite{7194592,8226297,chatzigeorgiou2014investigating}, and impact \cite{li2007empirical,5314231,cairo2018impact,bavota2015test} of code smells in software systems. Several recent works focus on domain-specific code smells including code smells from Machine Learning \cite{van2021prevalence,9680287,jebnoun2020scent},  Android \cite{10.1145/2897073.2897094, carvalho2019empirical}, and Data Intensive systems \cite{10.1145/3379597.3387467,bernard2021mongodb}. These works, a source of our inspiration, shed light on domain-specific issues relating to code smells and reveal their unique maintenance challenges. However, to the best of our knowledge, no prior work focuses on investigating the code smells in simulation modelling software.

Simulation modelling provides an imitative representation of real-world processes or systems in a controlled, virtual environment. It has found numerous applications in many critical areas including aviation, transportation, and medicine, which help their stakeholders with training, testing, and decision-making. Simulation modelling systems are also highly complex due to their extensive domain logic, which abstracts real, physical systems \cite{cellier2006continuous}. As a result, they could be susceptible to various quality issues including code smells. An investigation of code smells in simulation software systems can reveal important insights about their code quality issues and maintenance challenges. 

In this paper, we conduct an empirical study to investigate the prevalence, evolution, and impact of code smells found in simulation software systems. Our study relies on the analysis of 155 simulation and 327 traditional software systems collected from GitHub. We apply static analysis tools (e.g. Desginite \cite{designite}) to detect code smells in these systems. We also categorize the code smells into multiple categories according to their abstraction levels as well as Martin Fowler's catalog \cite{fowler2018refactoring} for our analysis. Through our experiments, we answer three important research questions as follows.

\begin{enumerate}[label=(\alph*)]
  \item \textbf{RQ1: Do simulation software systems smell like traditional software systems?}
  We use open-source static analysis tools (e.g. Designite \cite{designite}) and determine the prevalence of code smells in both simulation and traditional software systems. From our experiments, we find that there are significant differences in the distribution of code smells between simulation and traditional software systems. In particular, simulation systems contain 62.77\%, 19.1\%, and 26.7\% more \textit{Magic Number}, \textit{Long Statement}, and \textit{Long Parameter List} code smells respectively per line compared to those of traditional systems.  On the other hand, several smells such as \textit{Unutilized Abstraction}, \textit{Feature Concentration}, and \textit{Broken Modularization} are more prevalent in traditional software systems.
  
  \item \textbf{RQ2: How long do code smells last in simulation software systems?}
  We examine the evolution of code smells throughout the development stage of simulation software systems. Specifically, we inspect the probability and duration of survival of each type of code smell in simulation software systems and contrast them with their counterparts from traditional software systems. We observe that a typical code smell in a simulation system lasts 815 more days than its counterpart in traditional systems. Moreover, the \textit{Implementation smells} last the longest in simulation systems with a median survival time of 3,513 days. Our findings also show that most code-level problems are ignored during development.
  
  \item \textbf{RQ3: Do code smells co-occur with bugs in simulation software systems?}
  We mine both bug-inducing and bug-fixing commits from our collected systems using an open-source tool namely \emph{PyDriller} \cite{Spadini2018}, and examine the code smells from the mined commits. We conduct appropriate statistical tests (e.g. Chi-square coefficient and Cramer's V) to determine if code smells have any association with software bugs. We observed no significant association between the occurrence of code smells and bugs in simulation software systems.
  
\end{enumerate}

\textbf{Paper structure:} The rest of the paper is structured as follows. In Section \ref{sec:bg}, we outline background 
concepts related to
our study. Section \ref{sec:me} describes our research methodology. In Section \ref{sec:res}, we analyze the differences in the prevalence of code smells between simulation and traditional systems, the survivability of code smells in simulation and traditional systems, and the association between code smells and software bugs in simulation systems. The implications of our study are described in Section \ref{sec:imp}. Section \ref{sec:threat} focuses on the threats to validity, Section \ref{sec:rw} discusses related work, and finally, Section \ref{sec:con} concludes our paper.

\section{Background}
\label{sec:bg}
\subsection{Code Smells}
Code smells indicate a deeper design problem within a codebase. They are not bugs and do not prevent the software program from functioning correctly. However, they slow down the development process and increase the risk of software bugs or failures. Over the years, many static analysis tools (e.g. Designite, Understand, SonarLint) have been designed to detect code smells in software systems. Traditionally, code smells have been categorized into three groups according to their level of abstraction \cite{sharma2018survey}. We provide a brief outline of these code smells below.

\subsubsection{Implementation Smells} Implementation smells indicate issues in the code-level implementation of programming solutions. Examples of Implementation smells are as follows.
    \begin{itemize}
        \item \textbf{Magic Number} refers to hard-coded, unexplained numbers in the source code.
        \item \textbf{Long Statement} refers to a program statement that is too long and hard to understand.
        \item \textbf{Long Parameter List} refers to methods or constructors containing too many parameters (e.g. more than three).
        \item \textbf{Complex Method} refers to a method with high Cyclomatic complexity (e.g. more than 20) \cite{thomas2008software}. 
        \item \textbf{Empty Catch Clause} refers to try-catch blocks that do not properly handle an encountered error or exception. 
        \item \textbf{Long Method} Refers to a method containing too many lines of code.
        \item \textbf{Missing Default} refers to a missing default clause in a switch statement.
        \item \textbf{Long Identifier} refers to variables with unnecessarily long names.  
    \end{itemize}

\subsubsection{Design Smells} Design Smells indicate design and organization issues in the classes of a software system. Examples of Design smells are as follows.
    \begin{itemize}
        \item \textbf{Broken Hierarchy} refers to parent and child classes that do not share an \emph{IS-A} relationship. 
        \item \textbf{Broken Modularization} refers to multiple methods being scattered among different classes when they should be kept under a single class.
        \item \textbf{Cyclic Hierarchy} refers to the dependencies of a parent class on its child classes.
        \item \textbf{Deep Hierarchy} refers to excessively long hierarchical chains among the classes. 
        \item \textbf{Deficient Encapsulation} refers to classes that have greater access to other classes than what is required.
        \item \textbf{Feature Envy} refers to methods that are more interested in accessing data from other classes than from their class.
        \item \textbf{Unexploited Encapsulation} refers to client classes that use explicit type-checking via long if-else or switch chains instead of exploiting the \emph{polymorphism} principle.
        \item \textbf{Unutilized Abstraction} refers to an abstraction that is left unused or not directly used.
        \item \textbf{Wide Hierarchy} refers to inheritance hierarchies that have a large number of sub-types at the same level, leading to wide hierarchical structures. Due to the high number of sub-types, most sub-types might implement similar functionality. This might indicate missing intermediate types, which can be used to implement common functionalities across different sub-types. 
    \end{itemize}
    
\subsubsection{Architectural Smells} Architectural Smells indicate deeper problems within the architecture of a software system. Examples of Architectural smells are as follows.
    \begin{itemize}
        \item \textbf{Cyclic Dependency} refers to two or more abstractions that depend on each other. For example, in a system with components A and B, component A depends on a part of component B and component B depends on a part of component A. In this situation, components A and B are said to be cyclically dependent.
        \item \textbf{Dense Structure} refers to modules or packages with deep and complex hierarchies.
        \item \textbf{Feature Concentration} refers to components that implement more than one feature.
        \item \textbf{God Component} refers to a component or module that has too many classes and lines of code.
        \item \textbf{Unstable Dependency} refers to a component that depends upon another less stable component.
        \item \textbf{Scattered Functionality} refers to a system where multiple components are responsible for realizing the same high-level concern.
    \end{itemize}

Besides abstraction level, Fowler et al. \cite{fowler2018refactoring} also divide the code smells into five categories as follows.

\begin{itemize}
    \item \textbf{Bloaters} are methods, classes, and components that have increased to such a great extent that they are very hard to work with. Common examples include Long Statements, Long Methods, Long parameter lists, and Large Classes.
    \item \textbf{Object-Orientation Abusers} refer to code containing incorrect or incomplete applications of Object-Oriented principles. Code smells such as Switch Statements, Refused Bequest, and Unutilized Abstraction are common examples of Object-Oriented Abusers.
    \item \textbf{Change Preventers} refer to poorly designed classes and modules that prevent further code-level changes. Some examples of Change Preventers include Shotgun Surgery and Divergent Change.
    \item \textbf{Dispensables} refers to redundant parts of the code that can be removed without breaking existing functionality. Common examples include the Dead Code, Code Comments, Duplicate Code, Lazy Class, and Speculative Generality/Unutilized Abstraction smells.
    \item \textbf{Couplers} contribute to excessive coupling among classes. Some common examples of Couplers include Feature Envy and  Inappropriate Intimacy.
\end{itemize}

\subsection{Statistical Significance}
\label{sub-sec:sig}
The null hypothesis describes a situation when there is no relationship between two samples under analysis \cite{Helmenstine2010-lp}. In other words, if the hypothesis is true, then any apparent relationship can be left up to chance alone. A statistical significance test must be conducted to determine whether to reject or accept the null hypothesis.

Mann-Whitney U test \cite{10.1214/aoms/1177730491} is a non-parametric statistical test that compares two independent samples for statistical significance without assuming their underlying distribution. The test gives a $p$ value that measures the probability of the null hypothesis being true. The obtained $p$ values are usually contrasted against a predefined significance threshold $\alpha$, which is generally accepted as 0.05 \cite{neyman1976emergence}.

Cliff's delta \cite{cliff1993dominance} is another non-parametric statistical test that quantifies the differences between two samples and delivers a value between -1 and +1. The values closer to $+1$ indicate that all items in the first sample are higher than those of the second sample, while values closer to $-1$ indicate the opposite. On the other hand, values closer to 0 indicate little to no difference between the two samples. The Equation \ref{eq:cliff} shows that the outcome $\delta$ can be calculated using the number of items, $n1$, and $n2$, from two samples respectively. Here, $\delta_{ij}$ represents the number of times one sample's value exceeds the other.

\begin{equation}
    \delta = \frac{1}{n1 \times n2}  \sum\limits_{i=1}^{n1} \sum\limits_{j=1}^{n2} \delta_{ij}
    \label{eq:cliff}
\end{equation}

We use the Mann-Whitney U and Cliff's delta tests to answer our first research question. They show the differences in the prevalence of code smells between simulation and traditional software systems.

\subsection{Survivability Analysis}
Survivability analysis is a statistical technique that analyzes the expected delay of an event. An event can be anything that is clearly defined, such as death during a hospital stay or mechanical failure in a machine. To determine the probability of survival, events experienced by subjects within a given period must be tracked. If the subjects do not experience any event or leave before the event within the period of observation, then the event is censored at the end of the period.

Kaplan-Meier Estimator \cite{doi:10.1080/01621459.1958.10501452} is a survivability analysis technique that returns a probability of survival given a time to event and the status of a subject. The \emph{time to event} is defined as the interval between the start of observation and the occurrence of an event. The time interval can be measured in any positive unit.  On the other hand, \textit{status} is a boolean value that indicates whether a subject has experienced an event or the data is censored. \emph{The survivability function $S(t)$} returns the probability of survival of a subject after time $t$. As shown in Equation \ref{eq:Kaplan}, the \emph{survivability function $S(t)$} is calculated within the time when at least one event occurred ($t_i$). Here, $d_i$ is the number of events that occurred within the time $t_i$ and $n_i$ is the number of individuals that survived up to time $t_i$.

\begin{equation}
     S(t) = \prod_{i:t_i\leq t} [1 - \frac{d_i}{n_i}]\\
     \label{eq:Kaplan}
\end{equation}
 
We use the Kaplan-Meier Estimator to answer our second research question. We use its estimated probabilities of survival to determine which code smells last longer in simulation and traditional software systems.

\subsection{Contingency Table}
\label{sub-sec:cont}
Cross-tabulation or Contingency Tables are used to show the frequency distribution of multiple variables with a matrix representation. Each cell of a contingency table represents the Observed Frequency $O_i$ of a combination of a variable with another variable. Table \ref{table:contingency} shows an example of a contingency table for two categorical variables -- Smoker (Smoker and Non-Smoker) and Lung Cancer (Lung Cancer and No Lung Cancer). Here, we have 50 Smokers and 30 Non-Smokers with Lung cancer as well as 70 Smokers and 150 Non-Smokers without any Lung Cancer. These values show the frequency of both Smokers and Non-Smokers with and without Lung cancer, where each cell represents the Observed Frequency $O_i$ for both variables.

\begin{table}[htbp]
\centering
\caption{An Example of Contingency Table}
\label{table:contingency}
\begin{tabular}{|c||c|c|c|}
\hline
& \textbf{Lung Cancer} & \textbf{No Lung Cancer} & \textbf{Total} \\
\hline
\hline
\textbf{Smoker} & $50$ & $70$ & $120$ \\
\hline
\textbf{Non-Smoker} & $30$ & $150$ & $180$ \\
\hline
\textbf{Total} & $80$ & $220$ & $300$ \\
\hline
\end{tabular}
\end{table}

\subsection{Association Test}
\label{sub-sec:asst}
Association tests are used to determine the nature and strength of any associations between any two different variables. Pearson's Chi-Squared test \cite{pearson1900x} is one such statistical test that determines the association between two categorical variables. It compares the observed frequencies $O_i$ derived from a contingency table (see Section \ref{sub-sec:cont}) to the expected frequencies $E_i$. As described in Equation \ref{eq:expected_f}, we get different expected frequencies $E_i$ for each combination of variables. Each value $E_i$ shows the frequency of a combination assuming that both variables are independent. Here, $row$ and $col$ are the sum of frequencies in each row and column in a contingency table. By dividing the product of $row$ and $col$ with the overall number of samples ($n$) for all categories, we get our expected frequency $E_i$. In Equation \ref{eq:pearson}, we see that the Chi-squared statistic $\chi^2$ is derived from Observed frequency $O_i$ and expected frequency $E_i$ for each variable.

\begin{equation}
    E_i = \frac{row  \times  col}{n}
    \label{eq:expected_f}
\end{equation}

\begin{equation}
    \chi^2 = \sum \frac{(O_i - E_i)^2}{E_i}
    \label{eq:pearson}
\end{equation}

The Cramer's V \cite{cramer1999mathematical} test is an extension of Pearson's Chi-Squared test that measures the strength of association between two categorical variables. This test outputs a value between 0 and 1, where 0 indicates total independence and 1 indicates complete association. As outlined in Equation \ref{eq:cramer}, the association score $V$ can be derived from Pearson's Chi-square coefficient $\chi^2$. Here, $n$ is the total number of samples from all categories, while $row$ and $col$ represent the frequencies of two categorical values respectively.

\begin{equation}
    V = \sqrt{\frac{\chi^2}{n  \times  min(row-1, col-1)}}\\
    \label{eq:cramer}
\end{equation}

We use Pearson's Chi-squared and Cramer's V test to answer our third research question. While Pearson's Chi-squared test determines the presence of associations between software bugs and code smells, Cramer's V test is used to gauge the strength of their associations.

\begin{figure*}
    \hspace*{-1cm}
    \centering
    \includegraphics[scale=0.33]{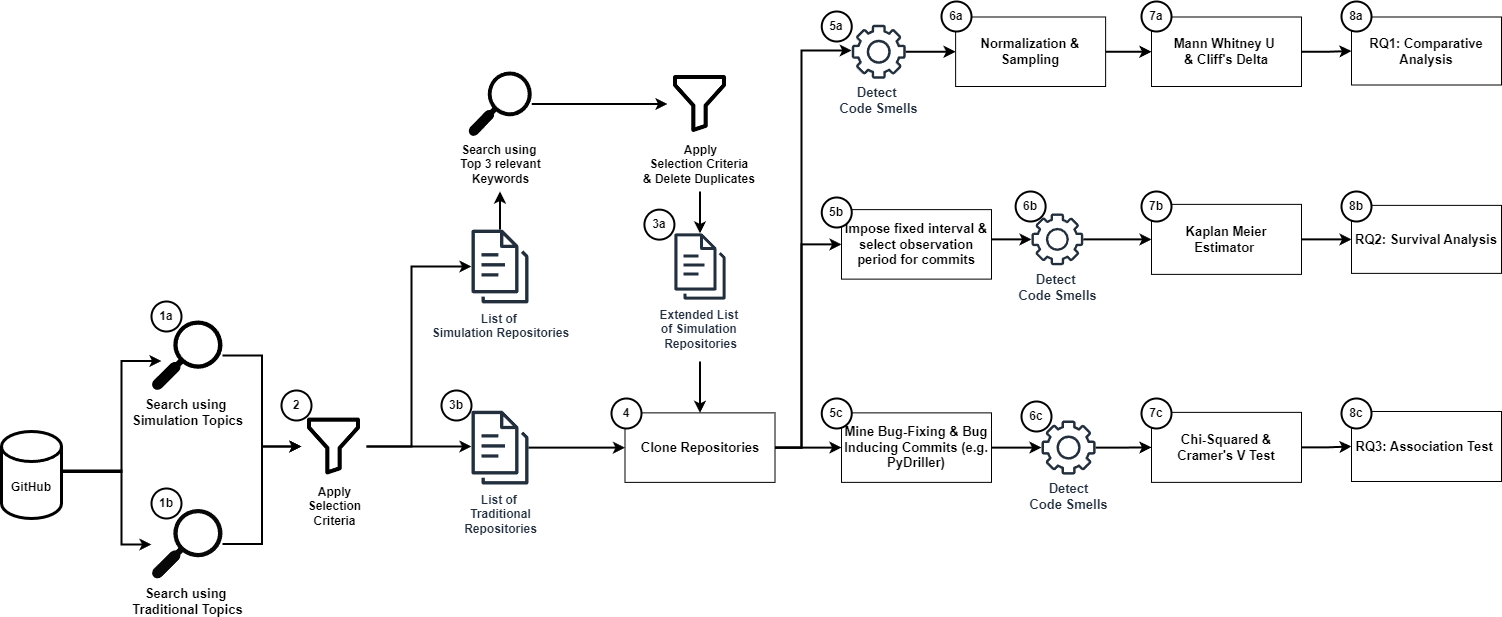}
    \caption{Schematic Diagram of our empirical study}
    \label{fig:process}
\end{figure*}

\section{Methodology}
\label{sec:me}
Figure \ref{fig:process} shows the schematic diagram of our conducted study. First, we select hundreds of repositories hosting simulation and traditional software systems from GitHub. After performing the necessary data preprocessing and cleanup, we detect code smells in the repositories using static analysis tools. Then, we perform a comparative study on the prevalence of code smells to answer our first research question. We also select commits from simulation software to perform a survivability analysis of the code smells. Finally, we attempt to find the association between code smells and bugs in the simulation systems. In the following sections, we discuss different steps of our methodology.

\subsection{Project Selection Criteria}
We use GitHub as a source of our simulation and traditional software systems. We only select the repositories that meet the following criteria:
\begin{enumerate}
        \item We only consider repositories with \emph{open-source licenses} (MIT, GPL, and Apache licenses) as they do not require explicit permissions for any third-party re-use.   
        \item We do not consider any repositories \emph{disabled} by the repository owner since we do not have permission to access them.
        \item We do not consider \emph{archived} repositories since they have stopped all active development.
        \item We do not consider any \emph{forks} of original repositories. This minimizes code duplication as forked projects share code with their original repository.
\end{enumerate}

\subsection{Data Collection}
As shown in Step 1a and 1b of Figure \ref{fig:process}, we use the GitHub search API \cite{githubGitHubREST} to collect repositories using GitHub Topics and keywords. First, we search using the following topics -- \emph{simulation}, \emph{simulator} and \emph{simulation modelling} --and collect a list of simulation systems. Similarly, we use several traditional topics -- \emph{web}, \emph{framework}, \emph{android}, \emph{HTTP}, and \emph{desktop} -- to collect a diverse set of traditional software systems. We also apply the selection criteria above (Step 2) and collect 67 simulation systems and 327 traditional systems.

Since we have significantly less number of simulation systems, we augment our existing list of simulation systems by following the process shown in Step 3a of Figure \ref{fig:process}. First, we retrieve the top three keywords (e.g. simulation, modelling, cloudsim) by analyzing the README documents of the retrieved simulation systems and using TI-IDF and RAKE scores \cite{rose2010automatic}. Then we perform a keyword search using these keywords to collect 775 new repositories. After applying our selection criteria above and manually analyzing repositories, we get a total of 88 new simulation systems, which leads to a total of 155 simulation systems. Finally,  as shown in Step 4, Figure \ref{fig:process}, we clone the final list of 155 simulation and 327 traditional systems for our analysis.

\subsection{Comparative Analysis}
As shown in Step 5a of Figure \ref{fig:process}, we start our comparative analysis by detecting code smells in simulation and traditional systems using the Designite tool \cite{designite}. It gives us a detailed report on code smells for each repository. Since repositories could have different sizes, to make the comparison fair, we normalize the number of smells in each repository against the total lines of code. This gives us an average number of smells from each repository, which accounts for repository size. After normalization, we perform a random sub-sampling on traditional software repositories (Step 5b, Figure \ref{fig:process}) to equalize the number of simulation modelling and traditional software systems. We then perform the \emph{Mann-Whitney U test} (see Section \ref{sub-sec:sig}) to determine if there are any significant differences in the distribution of smells between the simulation and traditional systems.

\subsection{Survivability Analysis}
\label{sub-sec:surv}
For our survivability analysis, we first collect all commits from each repository. Since we have 327 traditional systems against 155 simulation systems, we take a random sub-sample of 155 traditional systems for our analysis to reduce bias. Furthermore, since our selected simulation and traditional repositories contain thousands of commits, analyzing them all is computationally expensive. To remedy this, we use Sas et al's \cite{COMMIT} technique to select one commit every 4 weeks, which reduces the computational complexity. We then select the observation period by retrieving the earliest and latest commit times common to both simulation and traditional systems. This leaves us with an observation period of 4,949 days (Step 5b, Figure \ref{fig:process}).

After selecting our observation period, we ran the Designite tool on our selected commits to detect their code smells. This is done by using the \emph{multi-commit analysis} mode to analyze multiple selected commits at once \cite{designitetoolsCommandsx2014} (Step 5b, Figure \ref{fig:process}). Then we checked if any code smell was able to survive until the last commit by observing their probabilities of survival. To calculate the probability of survival of each smell, we use the Kaplan-Meier Estimator (see Section \ref{sub-sec:surv}). We collect the estimates for each type and sub-type of code smell across multiple repositories and determine their chances of survival during the development or evolution of a system.

\subsection{Mining Bug Inducing and Fixing Commits}
To determine any association between software bugs and code smells in simulation systems, their buggy and bug-fix commits should be analyzed. To this end, we use an appropriate list of keywords (e.g. Table \ref{table:contribution}) and find the bug-fix commits (Step 5c, Figure \ref{fig:process}). 
Table \ref{table:contribution} shows how each keyword helped us identify the bug-fix commits. These keywords were widely used by the existing works of Antoniol et al. \cite{antoniol2008bug}, Mockus et al. \cite{mockus2000identifying}, and Zhong et al. \cite{zhong2015empirical} to find bug-fix commits. We search for these keywords in the title of each commit to detect the bug-fix commits from each repository.

\begin{table}[]
\centering
\caption{Keywords for detecting bug-fix commits}
\begin{tabular}{|l|c|}
\hline
\multicolumn{1}{|l|}{\textbf{Keyword}} & \multicolumn{1}{l|}{\textbf{Detection of bug-fixing Commits}} \\ \hline
\hline
fix & 44.67\% \\ \hline
fixed & 24.19\% \\ \hline
bug & 9.35\% \\ \hline
issue & 4.18\% \\ \hline
except & 3.96\% \\ \hline
\end{tabular}
\label{table:contribution} 
\end{table}

After capturing the bug-fixing commits, we then use PyDriller \cite{Spadini2018} to find the corresponding bug-inducing commits. bug-inducing commits inject the bugs in the source documents that are later resolved by the bug-fixing commits. PyDriller relies on the SZZ algorithm \cite{sliwerski2005changes} for its operation.
The SZZ algorithm looks for potential bug-inducing commits by selecting bug-fixing commits and traversing through the version history to find commits that introduced changes in the currently fixed code.

\subsection{Association Test}
\label{sub-sec:me-as}
As shown in Step 6c of Figure \ref{fig:process}, we run Designite on both the bug-fixing and bug-inducing versions of code using the mined bug-fixing and bug-inducing commits. We use Designite's \emph{multi-commit analysis} mode to analyze both the bug-inducing and bug-fixing versions of the code to detect their code smells. After analyzing code smells in each bug-fixing and bug-inducing version, we organize the data into a contingency table with high and low frequencies of code smells for bug-inducing and bug-fixing categories (see Section \ref{sub-sec:cont}). Here, we discretize the frequencies using quantile binning to account for any imbalanced distribution of code smell frequencies \cite{zhang2020descriptive}. Then we calculate the Observed and Expected Frequencies $O_i$ and $E_i$ from our contingency table (see Section \ref{sub-sec:asst}) to find any associations between them. 

After getting our Observed and Expected Frequencies $O_i$ and $E_i$, we perform the Pearson's Chi-square test and  Cramer's V test respectively as outlined in Step 7c of Figure \ref{fig:process}. Here, Pearson's Chi-square test is used to determine the presence of an association between code smells and software bugs while Cramer's V test shows the strength of the association between code smells and buggy code.

\subsection{Replication Package}
We made our replication package \cite{Mahbub_On_the_Prevalence}  publicly available for any third-party reuse or replication.

\section{Results and Analysis}
\label{sec:res}

\subsection{ RQ1: Do simulation software systems smell like traditional software systems?}
\label{sub-sec:res-RQ1}
After collecting the code smell statistics from simulation and traditional systems, we compare their prevalence of code smells. In particular, we normalize the code smell frequencies against the lines of code (LOC) in each repository for our comparative analysis.

\begin{figure}
    \centering
    \includegraphics[scale=0.37]{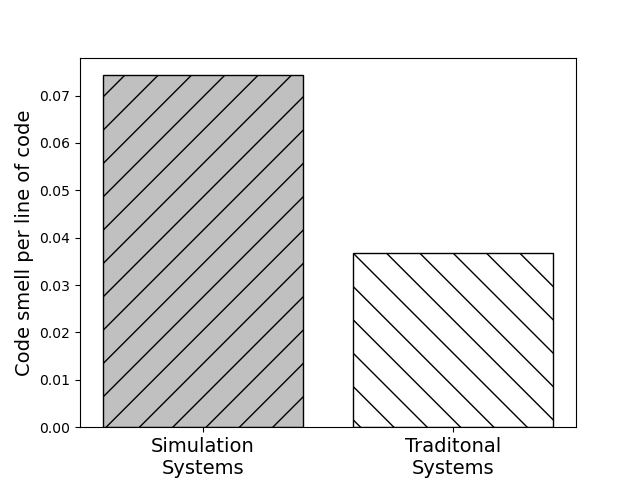}
    \caption{Prevalence of code smells in Simulation and Traditional systems}
    \label{fig:bar_comp}
\end{figure}

\begin{figure}
    \centering
    \includegraphics[scale=0.37]{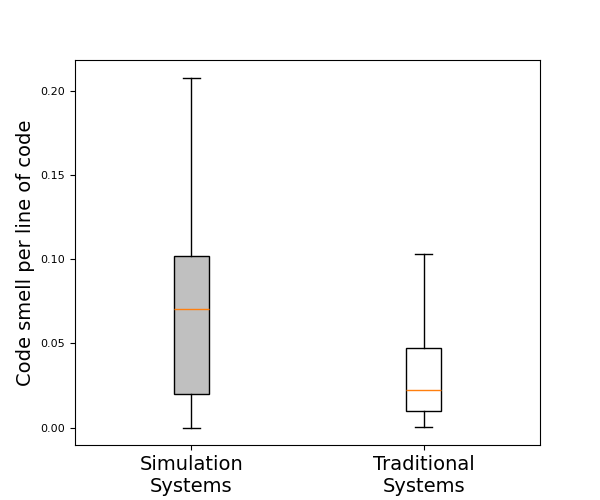}
    \caption{Distribution of normalized frequencies of code smells in Simulation and Traditional systems}
    \label{fig:box_comp}
\end{figure}

As shown in Figure \ref{fig:bar_comp}, simulation software systems have more code smells per line than traditional ones. It indicates that simulation systems are more prone to code smells, with a significantly higher median than their traditional counterparts (check Figure \ref{fig:box_comp}). Upon further analysis, we find that not only do simulation systems have more code smells, but also their distribution is different from that of traditional systems (Figure \ref{fig:box_comp}). This is supported by our non-parametric Mann-Whitney U test, which shows a p-value of $8.61 \times 10^{-12}$, denoting a statistically significant difference between the distribution of code smells. We also found Cliff's delta, $\delta=0.44$, suggesting that simulation systems have significantly more code smells per line than traditional systems with a \emph{medium} effect size.

\begin{figure}
    \centering
    \includegraphics[scale=0.47]{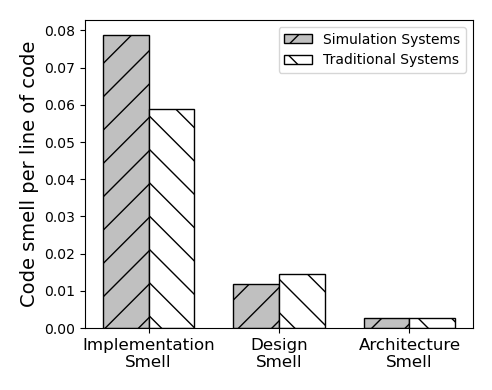}
    \caption{Prevelance of normalized frequencies of code smells in Simulation and Traditional systems (by smell type) }
    \label{fig:bar_type}
\end{figure}

\begin{figure}
    \centering
    \includegraphics[scale=0.43]{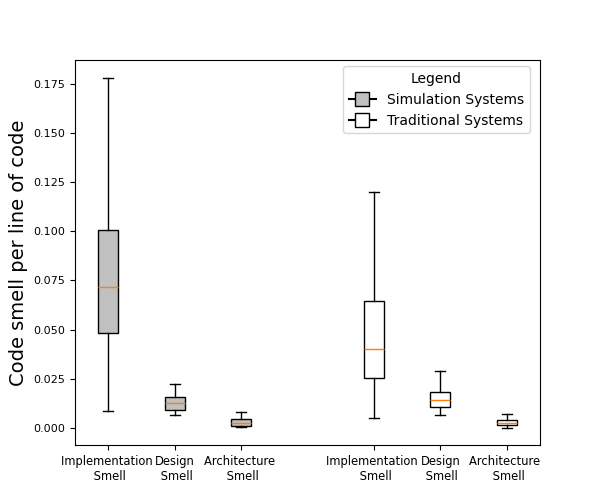}
    \caption{Distribution of normalized frequencies of code  in Simulation and Traditional systems (by smell type)}
    \label{fig:box_type}
\end{figure}

To achieve an in-depth insight, we analyze the prevalence of each category of code smells. As shown in Figure \ref{fig:bar_type}, simulation systems have a larger share of implementation code smells than their traditional counterparts. 
When frequency distribution is considered, Figure \ref{fig:box_type}, shows a 
bigger median
for Implementation code smells in simulation systems than that of traditional repositories. This observation is supported by a p-value of $4.05 \times 10^{-9}$ and Cliff's $\delta$ of $0.4569$. They suggest that the difference is statistically significant with simulation systems containing more implementation smells per line of code. Similarly, the distribution of Design smells is also different in both simulation and traditional repositories, as suggested by a p-value of $0.00634$ and Cliff's $\delta$ of $-0.1859$. Although this difference is much smaller, it still shows that simulation programs are home to fewer design smells than traditional repositories. Finally, the number and distribution of architectural smells in both repositories are similar. This is evidenced by the p-value of $0.6917$ and Cliff's delta of $-0.0311$, which suggest negligible differences between the prevalence of smells in both systems.

\begin{table}[]
\centering
\caption{Significance tests for code smell's prevalence}
\begin{tabular}{|l|c|c|}
\hline
\textbf{Code Smell} & \textbf{Cliffs $\delta$} & \textbf{p-value} \\ \hline
\hline
Long Parameter List & 0.1748 & 0.0231 \\ \hline
Long Statement & 0.1518 & 0.03703 \\ \hline
Magic Number & 0.5001 & $1.19 \times 10^-8$ \\ \hline
Empty Catch Clause & -0.4004 & $8.25 \times 10^-5$ \\ \hline
Unutilized Abstraction & -0.6384 & $3.24 \times 10^-9$ \\ \hline
Broken Modularization & -0.2783 & 0.0039 \\ \hline
Feature Concentration & -0.3284 & $6.59 \times 10^-6$ \\ \hline
\end{tabular}
\label{table:smell_stat}
\end{table}

We also compare the prevalence of each smell individually between the two types of systems. We employ Mann-Whitney-U, and Cliff's delta tests and Table \ref{table:smell_stat} summarizes our comparative analysis. From the 33 types of detected smells, only 7 were found to have a significantly different distribution $(p-value < 0.05)$. As shown in Table \ref{table:smell_stat}, the \textit{Long Parameter List}, \textit{Long Statement}, \textit{Magic Number}, and \textit{Empty Catch clause} belong to the Implementation Smell category, which confirms our findings in Figures \ref{fig:bar_type}, \ref{fig:box_type}. On the other hand, the \textit{Unutilized Abstraction} and \textit{Broken Modularization} smells belong to the \textit{Design Smell} category, whereas \textit{Feature Concentration} is a type of \textit{Architecture Smell}. We also see that the \textit{Long Parameter List}, \textit{Long Statement} and \textit{Magic Number} code smells have positive values of Cliff's delta. This indicates that these smells are more prevalent in simulation systems than in traditional systems. On the other hand, the Empty Catch Clause, Unutilized Abstraction, Broken Modularization, and Feature Concentration code smells all have negative values of Cliff's delta,  indicating their higher prevalence in traditional software systems.

\begin{figure}
    \centering
    \includegraphics[scale=0.37]{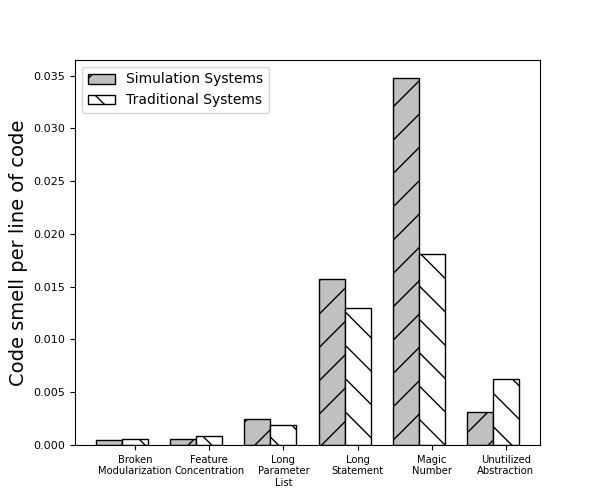}
    \caption{Prevalence of normalized frequencies of code smells in Simulation and Traditional systems (for significantly different code smells)}
    \label{fig:smell_different}
\end{figure}

From Figure \ref{fig:smell_different}, we see that the Magic Number and Long Statement smells occur 62.77\% and 26.72\% more times per line of code (LOC) in simulation systems, which supports our prior observation from Table \ref{table:smell_stat}. On the other hand, the Broken Modularization, Feature Concentration, and Unutilized Abstraction code smells occur 24.56\%, 38.22\%, and  66.42\% less times per LOC respectively. According to the above findings, simulation systems are more resistant to Design and Architectural code smells but significantly more vulnerable to implementation smells. Thus, while developing simulation software systems, developers should continuously refactor their implementation code to prevent the accumulation of technical debt throughout the system.

\begin{mdframed}[nobreak=true]
\textbf{Summary of RQ1:} We observe that code smells are more prevalent in simulation systems compared to traditional systems. This is especially true for \emph{Implementation Smells} such as \emph{Magic Number} and \emph{Long statement}, which occur much more frequently in simulation systems.
\end{mdframed}

\subsection{RQ2: How long do code smells last in simulation systems?}
\label{sub-sec:res-RQ2}
\begin{figure}
    \centering
    \includegraphics[scale=0.37]{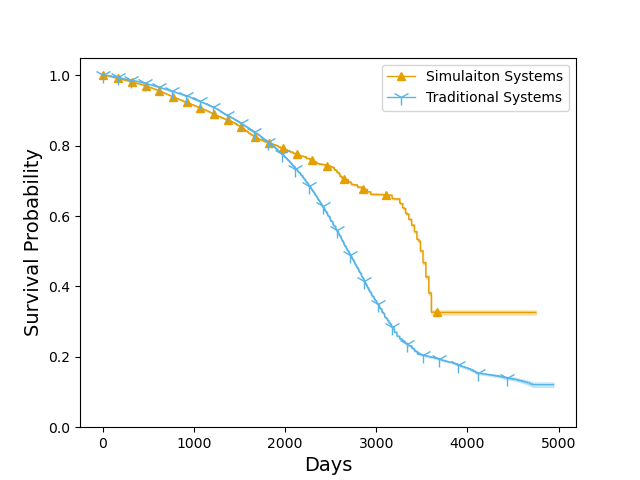}
    \caption{Survivability curves of code smells in simulation and traditional systems }
    \label{fig:kaplan_all}
\end{figure}
We perform our survivability analysis on 155 simulation and 327 traditional systems by first selecting commits with a 4-week interval. This leaves us with 2,263 and 5,402 commits from simulation and traditional systems respectively. Since we have more traditional systems, we use a random subsample of 155 traditional systems, leaving us with around 3007 traditional system commits. We select our observation period of 4,949 days by selecting common start and end dates between two systems (check Section \ref{sub-sec:surv}). Then we plot the Kaplan-Meier survivability curves for code smells detected in both simulation and traditional systems.

From Figure \ref{fig:kaplan_all}, we see that for the first 2,000 days, code smells in both simulation and traditional systems have similar probabilities of survival. However, after 2,000 days, their survivability curves start to diverge, with the probability of survival in traditional systems taking a nosedive. Although we observe a similar phenomenon in simulation systems after 3,000 days, their overall probability of survival remains higher when compared to traditional counterparts. At the end of the observation period, we observed that code smells in simulation and traditional systems have a median survival time of 3,513 and 2,698 days respectively. This shows that code smells can last longer in simulation systems than their counterparts in traditional systems.

\begin{figure}
    \centering
    \includegraphics[scale=0.37]{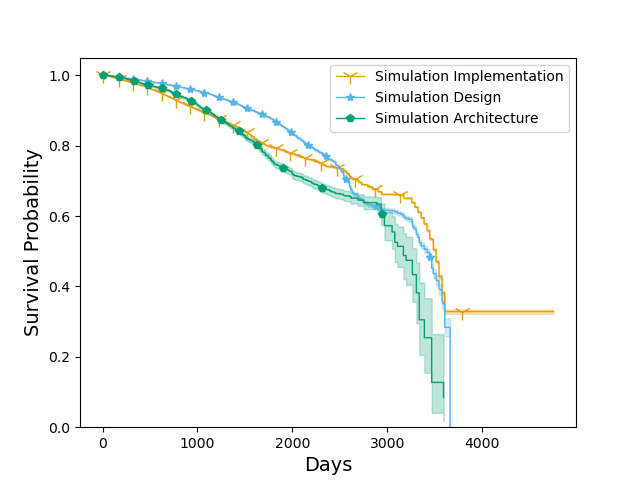}
    \caption{Survivability curves of code smells in Simulation systems (by abstraction level)}
    \label{fig:kaplan_break}
\end{figure}

Figure \ref{fig:kaplan_break} shows the survivability curves for three types of code smells --Implementation, Design, and Architecture smells -- from simulation systems. We notice that all share a similar probability of survival for around the first 3,000 days. However, after this, the probabilities of survival for \emph{Design} and \emph{Architecture Smells} decrease drastically, implying that most \emph{Design} and \emph{Architecture Smells} are refactored at the end of the observation period. However, the curve for implementation smells stays the same even after 3,500 days. This indicates that most implementation smells in simulation systems might not get refactored even after a long development time.

\begin{figure}
    \centering
    \includegraphics[scale=0.37]{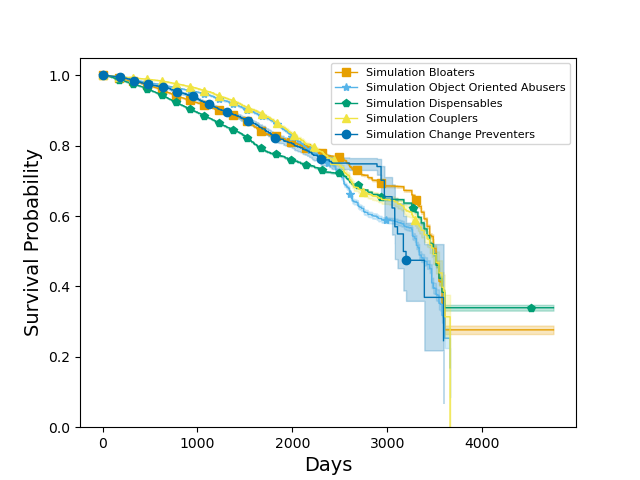}
    \caption{Survivability curves of code smells in Simulation systems (by smell type)}
    \label{fig:kaplan_change}
\end{figure}

We see a similar phenomenon in Figure \ref{fig:kaplan_change}, where all five types of code smells (i.e., Martin Fowler's catalog) have similar survivability curves for the first 2,500 days. After this, The \emph{Change Preventer} category of code smells shows a steeper survivability curve, ending with a 24.58\% chance of survival at the end of the observation period. These steeper curves suggest a longer duration of survival before refactoring. Similarly, the survivability curve of \emph{Object-Oriented Abusers} also diverges with a lower survival rate after 2,500 days, ending with a 16.85\% chance of survival at the end of the observation period. On the other hand, \emph{Bloaters}, \emph{Dispensables} and \emph{Couplers} share very similar survival curves up to the first 3,500 days. After this, we observe that the survivability curves for both \emph{Bloaters} and \emph{Dispensables} stagnate, having around 27.65\% and 33.93\% chance of survival respectively. Conversely, we see the survivability curve of \emph{Couplers} experience a sharp drop off after 3,500 days to almost 0\% chance of survival at the end of the observation period.


\begin{table}
\centering
\caption{Median Survival Times (MST) of all code smells}
\begin{tabular}{|l|c|}
\hline
\textbf{Code Smell} & \textbf{MST (days)} \\ \hline
\hline
Long Parameter List & 3,513\\ \hline
Long Statement & 3,467 \\ \hline
Complex Method & 3,513 \\\hline
Magic Number & 3,484\\\hline
Complex Conditional & 3,513 \\\hline
Long Method & 3,604 \\\hline
Empty catch clause & 3,322 \\\hline
Long Identifier & 3,576 \\\hline
Abstract Function Call From Constructor & 1,733 \\\hline
Unutilized Abstraction & 3,336 \\\hline
Deficient Encapsulation & 3,513\\\hline
Insufficient Modularization & 3,604 \\\hline
Broken Modularization & 3,107 \\\hline
Imperative Abstraction & 3,306 \\\hline
\textbf{Broken Hierarchy} & \textbf{3,661} \\\hline
Feature Envy & 3,390 \\\hline
Missing Hierarchy & 3,446 \\\hline
Unexploited Encapsulation & 1,809 \\\hline
Feature Concentration & 2,934 \\\hline
Unstable Dependency & 3,079 \\\hline
Scattered Functionality & 1,920 \\\hline
\end{tabular}

\label{table:median_time}
\end{table}

Table \ref{table:median_time} further shows the median survival times of all code smells detected in simulation software systems. Here, we see that the \textit{Broken Hierarchy} smell lasts the longest with a median survival time of 3,661 days followed by the \textit{Long Method} smell that has a survival time of 3,604 days. We also observe that most code smells last around 3,000 days in simulation systems. However, the \textit{Abstract Function Call From Constructor} and \textit{Scattered Functionality} are exceptions to this trend, lasting about 1,733 and 1,920 days respectively.

\begin{mdframed}[nobreak=true]
\textbf{Summary of RQ2:} We observe that the code smells in simulation systems can survive longer than the code smells in traditional systems. We also discover higher survival rates for \emph{Implementaion}, \emph{Bloaters} and \emph{Dispensable} code smells in simulation systems.
\end{mdframed}

\begin{figure}
    \centering
    \includegraphics[scale=0.43]{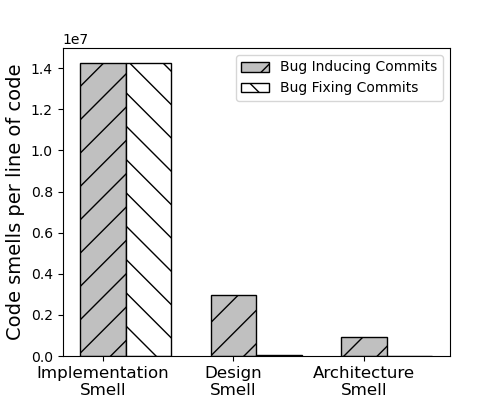}
    \caption{Number of code smells in bug-inducing and bug-fixing commits}
    \label{fig:bug-all}
\end{figure}

\subsection{RQ3: Do code smells co-occur with bugs in simulation software systems?}

We investigate whether code smells co-occur with software bugs by analyzing the prevalence of code smells in bug-inducing and bug-fixing commits of simulation systems. According to Figure \ref{fig:bug-all}, there are more \textit{Implementation Smells} than \textit{Design} and \textit{Architecture Smells} in both bug-inducing and bug-fixing commits. Interestingly, the implementation smells are equally present in both types of commits. This shows that either most \textit{Implementation smells} are introduced before bugs or they are not refactored even after bugs are fixed, implying that \emph{Implementation Smells} do not occur with bugs. This is backed up by our results from \emph{RQ2}, which shows that Implementation smells can survive the longest in simulation systems. However, as shown in Figure \ref{fig:bug-all}, bug-inducing commits contain a much larger amount of \textit{Design} and \textit{Architecture} smells than bug-fix commits. This shows that most \textit{Design} and \textit{Architecture Smells} are introduced at the same time as bugs in simulation systems.


\begin{table}[]
    \caption{Value of Cramer's V and Chi-square p values for each code smell}
    \centering
    \begin{tabular}{|l|c|c|}
        \hline
        \textbf{Code Smell} & \textbf{Cramer's V} & \textbf{p-value} \\ \hline
        \hline
        Imperative Abstraction & 0.196346 & 0.006165 \\ \hline
        Multifaceted Abstraction & 0.126884 & 0.389991 \\ \hline
        Feature Envy & 0.123369 & 0.081855 \\ \hline
        Broken Modularization & 0.113427 & 0.130604 \\ \hline
        Broken Hierarchy & 0.110897 & 0.158348 \\ \hline
        Feature Concentration & 0.102886 & 0.058798 \\ \hline
        Unstable Dependency & 0.091675 & 0.220785 \\ \hline
        Rebellious Hierarchy & 0.086085 & 0.446998 \\ \hline
        Scattered Functionality & 0.084107 & 0.788556 \\ \hline
        Missing Hierarchy & 0.082965 & 0.621711 \\ \hline
        Unutilized Abstraction & 0.075817 & 0.277980 \\ \hline
        Deficient Encapsulation & 0.058368 & 0.704499 \\ \hline
        Cyclic Hierarchy & 0.056553 & 0.884327 \\ \hline
        Unexploited Encapsulation & 0.045572 & 0.919234 \\ \hline
        Dense Structure & 0.045181 & 0.606460 \\ \hline
        Cyclic Dependency & 0.034300 & 0.942807 \\ \hline
        Insufficient Modularization & 0.031097 & 0.961491 \\ \hline
        Multipath Hierarchy & 0.029951 & 0.764551 \\ \hline
        Wide Hierarchy & 0.028609 & 0.628517 \\ \hline
        God Component & 0.005901 & 0.913614 \\ \hline
    \end{tabular}
    \label{table:cramers_v}
\end{table}
We also determine the association between code smells and bugs from a contingency table capturing observed and expected frequencies of each code smell (see Section \ref{sub-sec:me-as}). In particular, we perform Pearson's Chi-Squared and  Cramer's V tests to determine the association and strength of any associations between code smells and bugs. As we can see from Table \ref{table:cramers_v}, no single code smell has a p-value of <0.05, suggesting that code smells and bugs have no statistically significant association. Moreover, the value of Cramer's V for any smell does not cross 0.20, indicating a weak association between code smells and software bugs.

\begin{mdframed}[nobreak=true]
\textbf{Summary of RQ3:} We observe that \textit{Design} and \textit{Architecture} smells are introduced at the same time as bugs in the simulation system. However, we did not find any significant association between bugs and code smells in simulation systems.
\end{mdframed}

\section{Implication of Findings}
\label{sec:imp}
From \textit{RQ1}, we see that the prevalence of code smells in simulation systems is significantly different than in traditional systems. Two code smells -- \textit{Magic Number} and \textit{Long Statement} -- are 62.77\% and 26.72\% more frequent respectively in simulation systems. To avoid technical debt in simulation systems, developers should focus on adopting symbolic constants rather than \textit{Magic Numbers} in their code. Similarly, the Long Statements in code should be broken down into multiple manageable, smaller statements.

Our findings from \textit{RQ2} show that code smells in simulation software systems survive longer than those in traditional software systems. Smells such as \textit{Long Method}, and \textit{Broken Hierarchy} can survive 3,604 and 3,661 days respectively, on average, in a simulation system. This implies that these smells are possibly ignored during the development of simulation systems and have low priority during refactoring. From a subsequent analysis, we observe that \textit{Bloaters} and \textit{Dispensable} code smells are often ignored during development, as evidenced by their relatively high chances of survival - 33.93\% and 27.65\%. Thus, according to our findings, simulation systems could contain large complex code structures that are not properly maintained, leading to a significant amount of redundancy. To avoid accumulating technical debt, developers should prioritize refactoring these code smells in the simulation software systems.

Our \textit{RQ3} investigates the possible co-occurrences of bugs and code smells in simulation systems. Here, we found that \textit{Design} and \textit{Architecture Smells}  occur more frequently in the bug-inducing commits of simulation software systems. However, there is no significant association between code smells and bugs, as evidenced by our Pearson's Chi-Squared and Cramer's V tests. Thus, future investigations could focus on the impact of code smells in simulation systems and their refactoring strategies.

\section{Threats to Validity}
\label{sec:threat}
\textbf{Threats to internal validity.} Threats to \emph{internal validity} relate to any experimental errors or biases \cite{tian2014automated}. Since we select repositories from different domains, the quality of their code bases might not be comparable. To mitigate this threat, we used several metrics such as star count, number of contributors, and number of issues and attempted to select the repositories that are of comparable quality.  

We also use the SZZ algorithm to capture bug-inducing commits against their bug-fixing commits, where the algorithm has its limitations. Since we used a set of keywords to find bug-fixing commits, it may introduce false positives \cite{rodriguez2018reproducibility}. To remedy this, we manually checked 25 randomly sampled commits and found only 2 (8\%) false positive commits. Thus, any problems related to the SZZ algorithm might not significantly affect our overall findings. Besides, we have only considered differences, associations, or co-occurrences between two variables. As we do not claim any causation relationship between the two variables, the relevant threats might be minimal and might not affect our overall findings.

\textbf{Threats to conclusion validity.} Threats to \emph{conclusion validity} relate to the accuracy of conclusions \cite{garcia2012statistical}. During repository selection, we end up with 155 simulation systems and 327 traditional systems. Due to the differences in the sample sizes, the statistical results might be hard to infer. To mitigate this threat, we take a random subsample of equal size to make the comparison fair and unbiased.  We also use non-parametric statistical tests to avoid any assumptions behind the underlying distribution of the samples.

\textbf{Threats to External validity.} Threats to \emph{external validity} relate to the generalizability of any findings. To mitigate these threats and achieve diversity in our dataset,  we select our traditional systems from a variety of domains including web development, mobile development, and desktop application. Moreover, we collect simulation systems based on both topic-based and keyword-based searches. Thus, our selected systems might represent the general population of open-source simulation projects.

\section{Related Work}
\label{sec:rw}
\subsection{Prevalence of code smells}
Many existing works study the prevalence of code smells in different software systems. Sabóia et al. \cite{saboia2020prevalence} analyzed 25 C\# systems and suggest that \textit{Implementation smells} are the most prevalent in C\# systems. In particular, they found the \textit{Magic Number} and \textit{Long Statement} smells to be the most common in C\# systems. Similarly, Cardozo et al \cite{cardozo2023prevalence}  analyzed 24 Reinforcement Learning systems and found \textit{Long Method} and \textit{Long Method Chain} as the most common code smells. On the other hand, Jebnoun et al \cite{jebnoun2020scent} found no statistically significant difference in the prevalence of code smells between 59 deep learning and 59 traditional systems. Mannan et al \cite{10.1145/2897073.2897094} analyzed 500 Android and 750 Desktop applications and showed similar variety and densities of code smells for both systems. However, they observed that Desktop systems are dominated by \textit{External Duplication} and \textit{Internal Duplication} code smells, while Android systems contain an equal distribution of both smell types. In our study, we analyze the prevalence and distribution of code smells in 155 simulation software systems. Furthermore, we also measure the probabilities of survival for significantly different code smells from \emph{RQ1} for both simulation and traditional systems (see Section \ref{sub-sec:res-RQ1}). 

\subsection{Evolution of code smells}
Tufano et al \cite{7194592} found that very few code smells are introduced during software evolution. They also found 400 cases where refactoring operations introduced code smells in the system. Chatzigeorgiou et al \cite{chatzigeorgiou2014investigating} found that very few smells are removed from a system after their introduction. They observed that most code smells are removed through adaptive maintenance rather than active refactoring efforts. Muse et al \cite{10.1145/3379597.3387467} also analyzed 150 open-source Java projects and suggested that most SQL code smells can persist through multiple project versions without getting refactored. Unlike the above studies, we investigate the differences in the evolution of code smells from simulation and traditional systems.
We also find the survivability of each code smell in simulation systems across our observation period of 4,949 days (see Section \ref{sub-sec:res-RQ2}).

\subsection{Impact of code smells}
A literature survey of eighteen studies by Cairo et al \cite{cairo2018impact} found sixteen studies suggesting an association between bugs and code smells. The remaining two studies found no associations between bugs and code smells. According to Jaafar et al \cite{jaafar2017analysis}, Object-Oriented classes with anti-patterns and code clone smells are three times more likely to contain faults than non-smelly classes. In particular, up to 64\% classes contain co-occurrences of \textit{anti patterns} and \textit{Code Clone} smells, resulting in a higher fault proneness ratio. Hecht et al \cite{10.1145/2897073.2897100} found that refactoring of code smells can lead to memory and UI performance improvements in Android Systems. In particular, refactoring the \textit{Member Ignoring Method} smell leads to a 12.4\% improvement in UI performance. On the other hand, refactoring \textit{Garbage Collection} smells shows a 3.6\% improvement in memory performance. Our study attempts to detect
any associations between bugs and code smells in simulation systems. We also measure the strength of these associations to analyze the impact of code smells on bugs in simulation software.

\section{Conclusion}
\label{sec:con}
The presence of code smells in a software system indicates its deeper code quality issues. Thus, many studies focus on the prevalence and effects of code smells in various types of software systems. However, despite their enormous importance, there has not been any work on the code smells of simulation systems. In this study, we aim to fill this gap by analyzing 155 simulation and 327 traditional systems and investigating the prevalence, evolution, and impact of code smells in simulation modelling software. First, our analysis of simulation and traditional systems code smells shows that code smells are more prevalent in simulation systems compared to traditional systems. In particular, the \emph{Magic Number} and \emph{Long Statement} smells occur more frequently in simulation systems.  We also draw survivability curves to observe that code smells in simulation systems last longer than in traditional systems. Moreover, we find that \textit{Implementation Smells}, \textit{Bloaters}, and \textit{Dispensable} code smells have the highest survival rates in simulation systems. Lastly, after performing both Pearson's Chi-Square and Crammer's V test, we find no significant association between code smells and bugs. Overall, our study is one of the first to extensively investigate the nature of code smells in simulation software systems. By analyzing the prevalence, evolution, and impact of code smells in simulation modelling software, we shed light on the code quality of simulation systems. The results of our study can inform both users and developers of simulation software about specific threats to code quality and thus could impact their development practices. 

\section*{Acknowledgment}
This work was supported by the Climate Action Awareness Fund (CAAF) and Dalhousie University, Canada.




\balance
\bibliographystyle{IEEEtran}
\bibliography{references}

\end{document}